\documentstyle[LTpaper,psfig]{article}
\newcommand{\mywidth}{3.5in}
\begin{document}
%
%
\title{Boundary effects on the scaling of the superfluid density}
%
%
\author{N. Schultka$^a$ and E. Manousakis$^b$}
%
\address{\mbox{$^a$ Institut f\"ur Theoretische Physik, Technische Hochschule 
Aachen, D--52056 Aachen, Germany}\\
\vskip 10pt 
\mbox{$^b$ Department of Physics and Center for Materials Research and 
Technology, Florida State University,}\\
\mbox{Tallahassee, FL 32306, USA}}

%
\abstract{We study numerically the influence of the substrate 
(boundary conditions) on 
the finite--size scaling properties of the superfluid density $\rho_s$ 
in superfluid films of thickness $H$ within the XY model employing 
the Monte Carlo method.
Our results suggest that the jump $\rho_s H/T_c$ at the
Kosterlitz--Thouless 
transition temperature $T_c$ depends on the boundary conditions.
}
\maketitle\pagestyle{empty}

%
%

In Ref.\cite{SCHU1} we used the XY model to investigate numerically the 
finite--size scaling (FSS) properties of the superfluid density of superfluid
films with respect to the film thickness $H$. The superfluid density $\rho_s$
is related to the helicity modulus $\Upsilon$ of the XY model through 
$\rho_s=(m/\hbar)^2\Upsilon$ where $m$ denotes the mass of the helium
atom\cite{FISHER}. We used rather thin films up to $H=24$ and found that 
we were able to collapse our data for the helicity modulus onto a single
curve employing the scaling expression 
\begin{equation}
\frac{ \Upsilon(T,H)H_{eff} }{T}=\Phi(tH_{eff}^{1/\nu}), \label{sca}
\end{equation}
where the reduced temperature $t=T/T_{\lambda}-1$ and the effective thickness
$H_{eff}=H+D$ $(D=5.8)$. 
Applying this idea to the experimental data of Rhee et al. \cite{RHEE}
we achieved approximate data collapse as well $(D=0.145\mu m)$. Note that
the scaling form (\ref{sca}) reduces to the conventional
scaling form in the limit $H \gg D$. 
In Ref.\cite{SCHU1} we argued that the
increment $D$ could be understood as an effective scaling correction which
takes the influence of the boundary conditions (BC) into account, i.e.
vortex creation on a BC dependent length scale $l$. 

In order to investigate the role of the length scale $l$ further we use
the XY model on cubic lattices $L^3$ with periodic boundary conditions (PBC)
along the $x$-- and $y$--directions and Dirichlet boundary conditions (DBC)
(vanishing order parameter) along the $z$--direction and study the FSS behavior
of the $x$--component of the helicity modulus with respect to $L$. (In the 
following we will always refer to the $x$--component of the helicity modulus
as the helicity modulus.) By studying the FSS behavior of the helicity 
modulus of the XY model on cubic lattices with DBC in the $z$--direction we preserve the qualitative
features of films (suppression of the helicity modulus due to DBC 
(cf. also \cite{SCHU1})) but have the advantage of using ``thicker'' systems
(up to $L=48$). 

The hamiltonian of the XY model on a lattice is given by
${\cal H}=J\sum_{\langle i,j \rangle} \vec{s}_i \cdot \vec{s}_j$ where the
sum runs over nearest neighbors, the 
pseudospins $\vec{s}_i=(\cos\theta_i,\sin\theta_i)$ and sit on the 
lattice sites and $J$ sets the energy scale. The helicity modulus $\Upsilon$  
is defined as in Refs.\cite{TEITEL} and the temperature $T$ is measured in 
units of $J/k_{B}$. 
DBC are realized on different boundary inherent length scales as follows.
The layers $z=1,L$ are coupled to a spin configuration defined by
\begin{equation}
\vec{s}(x,y,z)=(-1)^{[\frac{x}{n}]+[\frac{y}{n}]} \, \vec{s}(1,1,z), \,\,\,\,
z=0,L+1, \label{stag}
\end{equation}
with $x,y=1,2,...,L$ denoting the integer coordinates of the lattice sites.
The symbol $[x/n]$ means the integer part of the number $x/n$.
The integer $n$ determines the linear dimension over which the spins of the
boundary spin configuration are parallel. Thus, the local magnetization in the 
boundary 
$\vec{M}=\sum_{i\epsilon A}\vec{s}_i$, with $A$ an area containing 
$2n\times 2n$ boundary spins, vanishes over the length scale $2n$, i.e.
the number $n$ serves as a measure for the length scale $l$ over which vortices
are created by the boundary. In our Monte Carlo simulations we used $n=1$ and 
$n=4$ and lattice sizes $L=24,32,40,48$.

\begin{figure}[htp]
\centerline{\psfig {figure=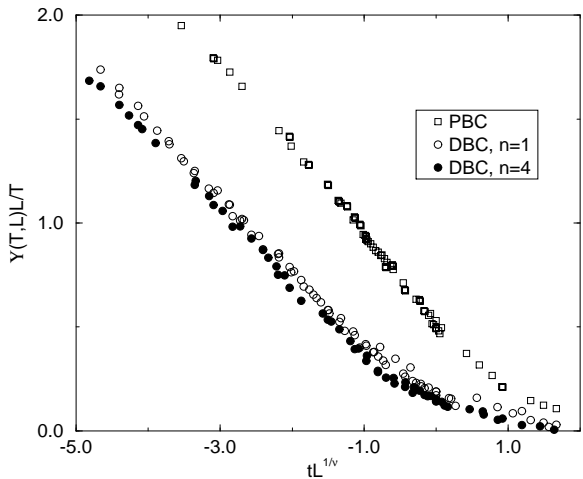,width=\mywidth}}
\caption{\label{fig1} The scaling function $\Upsilon(T,L)L/T$ for
PBC, DBC(n=1), and DBC(n=4) along the $z$--direction.}
\end{figure}

In Fig.\ref{fig1} we compare the FSS functions 
$\Upsilon(T,L)L/T=\Phi(tL^{1/\nu})$ ($\nu=0.6705$ \cite{GOLDNER}) for 
PBC, $n=1$ DBC, and $n=4$ DBC along the $z$--direction, respectively. We find
$\Phi(x)_{\mbox{PBC}} > \Phi(x)_{\mbox{DBC,n=1}} > \Phi(x)_{\mbox{DBC,n=4}}$.
This is qualitatively the same behavior as was demonstrated for the film 
case \cite{SCHU1}. To see the influence of the BC it is instructive to consider
the universal number $\Phi(0)$. We have $\Phi(0)=0.505(7)$ (PBC) 
\cite{SCHU3}, $\Phi(0)=0.189(2)$ (n=1 DBC), and $\Phi(0)=0.148(4)$ (n=4 DBC).
Combining these results and the results reported in Ref.\cite{SCHU1} 
we can immediately draw conclusions for the film case.
Due to the suppression of the helicity modulus we expect the 
Kosterlitz--Thouless transition temperatures $T_c(H)$ for films of 
thickness $H$ in the presence of DBC to satisfy the inequality 
$T_c^{n=1}(H) > T_c^{n=4}(H)$. Furthermore the inequality $D_{n=1}<D_{n=4}$ 
should hold for the increment $D$ which has to be added to the film thickness
$H$ to obtain $H_{eff}$. 
Thus, we suppose that the larger the length scale $l$ is (or $n$ in our
Monte Carlo calculations) the larger is the increment $D$ and the thicker
films have to be used to verify conventional FSS in experiments.

The dependence of the value $\Phi(0)$ on the BC suggests that in the case of
a film geometry the jump $\Upsilon(T_c(H),H)H/T_c(H)$ depend on the BC, too.
Table \ref{ta1} contains the values $\Upsilon(T_c(H),H)H_{eff}/T_c(H)$ for 
films with $n=1$ DBC for different film thicknesses and we
see that this value seems to saturate for $H\geq 16$ at about $0.97$. Thus,
for films with $H \gg D$ we would find $\Upsilon(T_c(H),H)H/T_c(H)=0.97$
compared to the value $2/\pi$ expected from renormalization group 
calculations\cite{NELSON} and found for films in the presence of PBC along
the $z$--direction\cite{SCHU4}  and in experiments\cite{BISHOP}.
If we applied $n=4$ DBC along the $z$--direction of the film we would expect
$\Upsilon(T_c(H),H)H/T_c(H)>0.97$.
\begin{table}[htp] \centering
 \begin{tabular}{|l|l|} 
 \hline
 \multicolumn{1}{|c|}{$H$}  & 
\multicolumn{1}{c|}{$\Upsilon(T_{c}^{2D}(H),H)H_{eff}/T_{c}^{2D}(H)$} \\
\hline
4  & 0.565(32) \\
8  & 0.81(17)  \\
12 & 0.870(65) \\
16 & 0.974(44) \\
20 & 0.972(68) \\
\hline
 \end{tabular}
\caption{\label{ta1} The jump
$\Upsilon(T_{c}^{2D}(H),H)H_{eff}/T_{c}^{2D}(H)$ for different 
thicknesses $H$.}
\end{table}

At this point we find it quite interesting to investigate the FSS behavior
of the superfluid density of superfluid films over a wide range of 
film thicknesses and for various substrates 
so that BC are realized which create vortices on different
length scales. These experiments could directly test our ideas of scaling
with an effective thickness for rather thin films and the dependence of the
jump $\Upsilon(T_c(H),H)H/T_c(H)$ on the BC. It would be especially interesting
to check these ideas using silicon as a substrate and the experimental
set-up of the experiment of Rhee et al.\cite{RHEE}.
%
%

\end{document}